\documentclass[12pt]{iopart}

\usepackage{graphicx}
\usepackage{epstopdf}
\usepackage{dcolumn}
\usepackage{bm}
\usepackage{rotating}

\begin{document}

\noindent{\it PREPRINT submitted to Journal of Physics B}

\title{Electronic structure of the Magnesium hydride molecular ion}

\author{M Aymar}
\address{Laboratoire Aim\'{e} Cotton, CNRS, B\^at. 505, Univ Paris-Sud,\\F-91405 Orsay Cedex, France}
\ead{mireille.aymar@lac.u-psud.fr}

\author{R Gu\'erout}
\address{Laboratoire Aim\'{e} Cotton, CNRS, B\^at. 505, Univ Paris-Sud,\\F-91405 Orsay Cedex, France}
\ead{romain.guerout@lac.u-psud.fr}

\author{M Sahlaoui}
\address{Laboratoire Aim\'{e} Cotton, CNRS, B\^at. 505, Univ Paris-Sud,\\F-91405 Orsay Cedex, France}
\ead{m\_sahlaoui@mail.univ-tlemcen.dz}

\author{O Dulieu}
\address{Laboratoire Aim\'{e} Cotton, CNRS, B\^at. 505, Univ Paris-Sud,\\F-91405 Orsay Cedex, France}
\ead{olivier.dulieu@lac.u-psud.fr}

\begin{abstract}
In this paper, using a standard quantum chemistry approach based on pseudopotentials for atomic core representation, Gaussian basis sets, and effective  core polarization  potentials, we investigate the electronic properties of the MgH$^+$ ion. We first determine potential energy curves for several states using different basis sets and discuss their predicted accuracy by comparing our values of the well depths and position with other available results. We then calculate permanent and transition dipole moments for several transitions. Finally  for the first time, we calculate the static dipole polarizability of MgH$^+$ as function of the interatomic distance. This study represents the first step towards the modeling of collisions between trapped cold Mg$^+$ ions and H$_2$ molecules.
\end{abstract}

\maketitle




\section{Introduction}

The reactivity of molecular systems in the gas phase at low temperature (as low as about 10~K) is of crucial importance for the evolution of cold natural environments in the interstellar medium. In the laboratory, amazing progresses have been achieved in the obtention of gaseous molecular samples at even lower temperatures, from the cold regime (a few K or less) down to the ultracold regime (a few mK and below). It is now possible to study chemical reactivity in the Kelvin range down to 6 orders of magnitude in temperature, with a wide variety of molecular systems \cite{smith2008}. In this context, molecular ions are appealing as they can be easily manipulated by electromagnetic fields to guide and confine them in traps, until the formation of ion crystals \cite{gerlich2008a}. Nowadays, cold molecular ions can be produced by sympathetic cooling with laser-cooled ions \cite{gerlich2008b} or by collisions with a buffer gas of helium of hydrogen \cite{gerlich2009}. Interactions between cold atoms and ions are also relevant for open question related to molecule formation in Bose-Einstein condensates \cite{cote2002} and to quantum information \cite{idziaszek2007}.

The first evidence for the formation of molecular ions in a laser-cooled ion trap has been reported by Baba and Waki \cite{baba1996}, by introducing air in a Mg$^+$ trap. A more controlled experiment has been achieved by Drewsen and coworkers \cite{molhave2000}, who introduced thermal H$_2$ or D$_2$ gas in a laser-cooled Mg$^+$ trap yielding trapped MgH$^+$ or MgD$^+$ ions. Other diatomic or triatomic molecular species have been produced, which are for instance reviewed in ref.\cite{roth2006a}. The main challenge is the detection of the formed molecular species, which is most often indirect, but non-destructive. Indeed, the molecular ions are not sensitive to the cooling laser, so they do not contribute to the fluorescence of the trap. By controlling the gaseous species which is introduced in the cold ion trap, and knowing the open reactive channels, one can observe the absence of fluorescence at specific locations of the ion crystals, assigned to the presence of molecular ions. The changes in the spatial structure and the excitation by an oscillating voltage \cite{drewsen2004} of the kinetic modes of the dual-species ion crystal can also be detected, and modeled considering the masses of the expected molecular ions. A more direct detection method - probably destructive- is needed, relying on the spectroscopic characteristics of the molecular ions. This would also allow to address the other major challenge of such experiments, i.e. the knowledge and the control of the internal energy of the formed molecular ions after the reactive process.

In this perspective, the MgH$^+$ ion represents a prototype system, which is extensively studied by the Aarhus group. Bertelsen {\it et al} \cite{bertelsen2004} observed for the first time the photodissociation of trapped MgH$^+$ ions embedded in a Mg$^+$ ion crystal, using a two-photon resonant scheme. The simulation of the MgH$^+$ photodissociation and the possibility to control the branching ratio has been investigated by J\o rgensen  {\it et al} \cite{jorgensen2005}. The isotope effects in the reaction of excited Mg$^+$ ions with either H$_2$ of HD molecules have been observed \cite{staanum2008} through a factor of 5 difference in the branching ratios for MgH$^+$ and MgD$^+$ ions.

It is also worthwhile to mention that prior to the above works, the effective two-electron nature of the MgH$^+$ molecule made it appropriate as a prototype for accurate quantum chemistry calculations at various stages of development of computing facilities \cite{numrich1975,olson1979,garcia-madronal1992,dutta1997}. The MgH$^+$ system also attracted attention in various situations of collision physics. Charge transfer in collision between H$^+$ and Mg have been investigated in the keV energy range \cite{morgan1979,dutta1997}, and related cross sections have been calculated \cite{olson1979}. Broadening of the Mg$^+$ resonance line by collision with atomic hydrogen has been investigated by Monteiro {\it et al} \cite{monteiro1988}. Dalleska {\it et al} investigated the energy dependence of the Mg$^+$+H$_2$ reactive collision and estimated the MgH$^+$ dissociation energy \cite{dalleska1993}. In most of these works, spectroscopic constant for various states of MgH$^+$ have been reported. Nevertheless, to our knowledge, there is only one high-resolution spectroscopic study of MgH$^+$, performed by Balfour in 1972 \cite{balfour1972}. The rotational analysis of the $A^1\Sigma^+ \rightarrow X^1\Sigma^+$ and $B^1\Pi \rightarrow X^1\Sigma^+$ transitions of several isopotomers ($^{24}$MgH$^+$, $^{25}$MgH$^+$, $^{26}$MgH$^+$) enabled the author to derive the low part of the potential curves for the $X$ and the $A$ states, and some spectroscopic constants for the $B$ state. According to ref.\cite{numrich1975}, another spectroscopic analysis has been performed by Numrich \cite{numrich1974}, but is not available to us.

In this work, we investigate electronic properties of MgH$^+$ ion as a first step towards the modeling of collisions between trapped cold Mg$^+$ ions and H$_2$ molecules. Following our work on effective two-electron diatomic molecules like alkali dimers \cite{aymar2005,aymar2006,deiglmayr2008} and alkali hydrides \cite{aymar2009}, we determine potential curves for various symmetries up to the Mg$^+(5s)$+H(n=1) asymptote, static dipole polarizabilities, and permanent and transition dipole moments for the main transitions involving the lowest electronic states relevant for photodissociation. In several occurrences, such results are given for the first time. We will often use atomic units for distances ($a_0$=0.052917720859~nm), energies ($2R_{\infty}=219474.63137054$~cm$^{-1}$), and dipole moments (1a.u.=2.54174~Debye).

\section{Computational approach}
\label{sec:method}

As in our previous work on alkali dimers \cite{aymar2005,aymar2006,deiglmayr2008} or alkali hydrides \cite{aymar2009}, we use an automated procedure to run the CIPSI package (Configuration Interaction by Perturbation of a multiconfiguration wave function Selected Iteratively) \cite{huron1973}.  The approach is based on the  gaussian $\ell$-dependent effective core potential (ECP) of Durand and Barthelat \cite{durand1974,durand1975} to represent the ten-electron Mg$^{2+}$ ionic core, Gaussian basis sets, and $\ell$-dependent effective terms for core polarization (CPP) along the lines of ref.\cite{muller1984} revisited by Foucrault {\it et al} \cite{foucrault1992}. The MgH$^+$ molecule is therefore modeled as an effective two-electron system. The molecular  orbitals are determined by restricted  Hartree-Fock  single electron calculations, including the CPP \cite{foucrault1992} providing the potential curves for the relevant molecular cations. Then a full configuration interaction (FCI) is  performed for each relevant molecular symmetry, providing potential curves, and permanent and transition dipole moments.

We will compare our results to those obtained by other various theoretical methods falling in two categories:

\begin{itemize}
\item {\it Effective core potential methods}: Fuentealba  and Reyes \cite{fuentealba1987} investigated the ground state of alkaline-earth monohydrides ions using a ECP+CPP approach similar to ours, and local spin density functional for the valence correlation, instead of a CI. In the context of the charge transfer between Mg and H$^+$ at low-keV energies, Dutta {\it et al} \cite{dutta1997} determined numerous MgH$^+$ potential curves and couplings, using a slightly different $\ell$-dependent gaussian ECP including polarization terms, a basis of Slater-type orbitals, and a partial CI. A similar work using a simple exponential Hellman-type ECP was previously reported by Numrich  and Truhlar \cite{numrich1975}. Finally, Garc\'ia-Madron\~nal {\it et al} \cite{garcia-madronal1992} performed FCI calculations based on an ECP usually referred to as model potential for the Mg$^+$ ion.

\item {\it All-electron calculations}: The most recent all-electron calculations have been reported in refs.\cite{vogelius2005,jorgensen2005}, improving the previous similar computations by Monteiro {\it et al} \cite{monteiro1988}. The authors employed a multireference configuration interaction (MRCI) approach to deliver adiabatic potential curves and dipole moments at the multiconfiguration self-consistent field level of the theory (MC-SCF), within the commercial MOLPRO package. Prior to this work, Rosmus and Meyer \cite{rosmus1977} studied the MgH$^+$ ground state with the coupled-electron pair approach (CEPA) and CI based on pseudonatural orbitals. Olson and Liu \cite{olson1979} performed SCF and CI calculations with Slater-type-function basis set with the ALCHEMY package. Spectroscopic constants of the MgH$^+$ ground state  have been calculated by Canuto {\it et al} \cite{canuto1993}, using fourth-order-many-body perturbation approach and the GAUSSIAN package.
\end{itemize}

As a preliminary step, we checked the convergence of our calculations as a function of the size of the Gaussian basis sets describing H and Mg$^+$.

For hydrogen atom, we considered three basis sets available in the literature. The set \{H$^{(A)}$\} corresponds to the large uncontracted basis \{$10s5p2d$\}  that we used for alkali hydrides \cite{aymar2009}. The  basis set \{H$^{(B)}$\} (\{$9s5p1d$\} contracted to \{$7sp5p1d$\}) has been used to study the electronic structure of HCl$^-$ \cite{oneil1986}, and BaH \cite{allouche1992}. The smaller basis set \{H$^{(C)}$\} (\{$7s3p2d$\} contracted to \{$6s3p2d$\}) has been used in ref.\cite{geum2001} to investigate the structure of alkali hydrides. The deviations for the calculated hydrogen ground state binding energy, compared to the experimental one (taken at the non-relativistic value of -0.5~a.u.), are $\Delta_{1s}$ (cm$^{-1}$)=3.1, 16, 64, for the three basis sets \{H$^{(A)}$\}, \{H$^{(B)}$\}, \{H$^{(C)}$\}, respectively.

The basis sets for Mg$^+$ has been chosen (Table \ref{tab:basis}) in combination with the adjustment of the cut-off radii $\rho_s$, $\rho_p$, $\rho_d$, $\rho_f$ of the CPP term, to match the experimental energy of the lowest levels of the ion, or their spin-orbit averaged energy, when appropriate (Table \ref{tab:atom}). We first used the basis set \{Mg$^{(A)}$\}, proposed in ref. \cite{durand1989} (\{$11s6p1d$\} contracted in \{$4s3p1d$\}) in the study of the structure and the stability of Mg$^+_n$ and Mg$^{++}_n$ clusters. The lowest $d$ level of Mg$^+$ cannot be reproduced accurately with this basis, which is due to the lack of $d$ orbitals. Therefore we extended the basis of ref. \cite{durand1987} (\{$4s2p1d$\} contracted to \{$3s2p1d$\}) towards the sets \{Mg$^{(B)}$\} (\{$6s4p3d$\} contracted to \{$5s4p2d$\}) and \{Mg$^{(C)}$)\} (\{$7s5p3d2f$\} contracted to \{$6s5p2d2f$\}), which are able to reproduce Mg$^+$ experimental binding energies $3s$, $3p$ and $3d$ exactly, and those of the higher excited levels $4s$, $4p$, $5s$ at about 1.5\% or better. This agreement is similar to the one obtained for alkali atoms in our previous works.

The comparison with the results of refs.\cite{garcia-madronal1992,vogelius2005,jorgensen2005} can be carried out from the point of view of binding energies, or energy differences (i.e. excitation energies, when selection rules apply). We first extracted from ref.\cite{garcia-madronal1992} atomic binding energies from the value of potential energies at $R=30a_0$. We see that due to their simple form of their model potential involving only two parameters, the authors cannot match the binding energy of the Mg$^+$ ground level. Moreover they did not include $d$ orbitals in their basis, so that their $3d$ level is indeed ill-defined, as it is visible in Table \ref{tab:atom}. Their lowest energy differences are in better agreement with experiment (again, except the one for the $3s-3d$ difference) than binding energies, which suggests that their model potential leads to a global upward energy shift of the excitation spectrum. Refs.\cite{vogelius2005,jorgensen2005} report all-electron calculations which cannot provide a precise binding energy for the ground state, so that the authors simply shifted the whole atomic spectrum to match the experimental ground level energy. Thus they obtained an excitation energy for the $3s-3p$ transition in reasonable agreement (at a 4\% level) with experiment.

\begin{table}[t]
\begin{tabular} {|c|c|c|c|c|c|c|}\hline
&\multicolumn{6}{|c|}{Basis sets} \\
&\multicolumn{2}{|c|}{\{Mg$^{(A)}$)\}}&\multicolumn{2}{|c|}{\{Mg$^{(B)}$\}}&\multicolumn{2}{|c|}{\{Mg$^{(C)}$\}} \\
&\multicolumn{2}{|c|}{\{$11s6p1d$\}/\{$4s3p1d$\}}& \multicolumn{2}{|c|}{\{$5s4p4d$\}/\{$4s3p3d$\}}& \multicolumn{2}{|c|}{\{$7s5p3d2f$\}/\{$6s5p2d2f$\}} \\ \hline
Orbitals&(a)&(b)&(a)&(b)&(a)&(b) \\ \hline
\{$s$\}&4917.3643 &-0.0058158 &4.302235&-0.013469&4.302235&-0.013469\\
       &739.76679 &-0.04345243&0.753995&-0.134665&0.753995&-0.134665\\
       &168.15033 &-0.1916909 &        &         &        & \\
       &47.059787 &-0.483461  &0.101389&1        &0.067089&1\\
       &14.395288 &-0.4198837 &0.037232&1        &0.0390  &1\\
       &          &           &0.016674&1        &0.01752 &1\\
       &          &           &0.00543 &1        &0.006   &1\\
       &24.982709 &0.0895159  &        &         &0.003   &1\\
       &2.458504  &-0.5718344 &        &         &0.001   &1\\
       &0.83418392&-0.5065272 &        &         &        & \\
       &          &           &        &         &        &\\
       &0.91772568&0.1277313  &        &         &        & \\
       &0.10346384&-0.6320582 &        &         &        & \\
       &          &           &        &         &        & \\
       &0.03809126&-0.455333  &        &         &        & \\  \hline
\{$p$\}&68.624283&0.0256489 &0.2  &1&0.2   &1\\
       &15.456213&0.1569184 &0.085&1&0.085 &1\\
       &4.5014417&0.4265914 &0.020&1&0.020 &1\\
       &1.3612957&0.5140575 &0.004&1&0.004 &1\\
       &         &          &     & &0.0009&1\\
       &0.36306322&0.1203396&     & &      & \\
       &          &         &     & &      & \\
       &0.086     &1.       &     & &      & \\ \hline
\{$d$\}&0.2&1.    &1.4553  &0.02754&1.455   &0.02754 \\
       &   &      &0.4332  &0.05391&0.4332  &0.05391 \\
       &   &      &        &       &        &\\
       &   &      &0.10839 &1      &0.10839 &1 \\
       &   &      &0.024635&1      &0.024635&1 \\ \hline
\{$f$\}& & & & &0.015 &1 \\
       & & & & &0.0005&1 \\ \hline
$\alpha$&\multicolumn{2}{|c|}{0.46904}&\multicolumn{2}{|c|}{0.46904}&\multicolumn{2}{|c|}{0.46904} \\
$\rho_s$, $\rho_p$,$\rho_d$, $\rho_f$ &\multicolumn{2}{|c|}{1.2397400,0.8966,1.5,-}& \multicolumn{2}{|c|}{1.3199, 1.253645, 1.5, -}& \multicolumn{2}{|c|}{0.9,1.25499,1.5,0.4}\\ \hline
\end{tabular}
\caption{Exponents (a) and contraction coefficients (b) of the gaussian orbitals of the various basis sets presented in the text for Mg$^+$, static polarizability $\alpha$ (in a.u.) of Mg$^+$ and adjusted cut-off radii $\rho_s$, $\rho_p$, $\rho_d$, $\rho_f$ (in a.u.) present in the $\ell$-dependent CPP term. The sets of orbitals which are contracted together are separated by a blank line.}
\label{tab:basis}
\end{table}

\begin{table}[t]
\begin{tabular} {|c|c|c|c|c|c|c|c|c|c|}\hline
Asymptote&\multicolumn{3}{|c|}{Experiment}&\multicolumn{3}{|c|}{$E_{calc}-E_{exp}$(cm$^{-1}$)}& \multicolumn{2}{|c|}{ref.\cite{garcia-madronal1992}(cm$^{-1}$)}&ref.\cite{jorgensen2005} \\
	         &(I)(a.u.)&(II)(cm$^{-1}$)&(III)(cm$^{-1}$)&(a)&(b)&(c)& (d)& (e)& (f)(cm$^{-1}$) \\
Mg$^+(3s)$+H    &-1.052536&-231004.9&0      &0   &0   &0   &557  &0   &0 \\
Mg$^+(3p)$+H    &-0.889736&-195274.6&35730.4&0   &0   &0   &710  &154 &-1230 \\
Mg$(3s^2)$+H$^+$&-0.833530&-182938.7&48066.3&733 &605 &-444&3650 &3093&1197 \\
Mg$^+(4s)$+H    &-0.734481&-161200.0&69805.0&4258&-779&-177&523  &-34 &	- \\
Mg$^+(3d)$+H    &-0.726801&-159514.4&71490.5&-   &0   &0   &10052&9496&	- \\
Mg$^+(4p)$+H    &-0.685113&-150364.9&80640.0&-   &-206&-205&1100 &544 &	- \\ \hline
\end{tabular}
\caption {Energy differences $E_{calc}-E_{exp}$ (in cm$^{-1}$) between calculated and experimental values of the lowest Mg$^+$+H($n=1$) asymptotes, and of the Mg($3s^2$)+H$^+$ asymptote, obtained with the various basis sets presented in the text. Experimental values are given for binding energies (columns (I) and (II)) and for excitation energies (column (III)). Columns (a), (b), (c), hold for the results using basis sets \{Mg$^{(A)}$)\}, \{Mg$^{(B)}$\}, and \{Mg$^{(C)}$\}, respectively. In column (d), energy differences are deduced from the values of the potential curves computed in ref.\cite{garcia-madronal1992} at $R=30$a.u., and from the corresponding excitation energies (column (e)). In column (f), energy differences are extracted from the excitation energies reported in ref.\cite{vogelius2005,jorgensen2005}.}
\label{tab:atom}
\end{table}

Prior to molecular calculations, we first performed a FCI for the neutral Mg atom, yielding ground state energies reported in Table \ref{tab:atom}. We obtain a reasonable agreement with the experimental value, comparable to the one obtained for alkali negative ions in our previous work \cite{aymar2006}. The largest basis set gives a ground state energy lower than the experimental one, which may be partly due to negative energy differences obtained for the highly-excited levels in Mg$^+$. Our value is in better agreement than the one obtained in the other available works.




\section{Potential curve calculations}
\label{sec:pot}

\begin{figure}[htb]
\begin{center}
\includegraphics[width=0.8\textwidth]{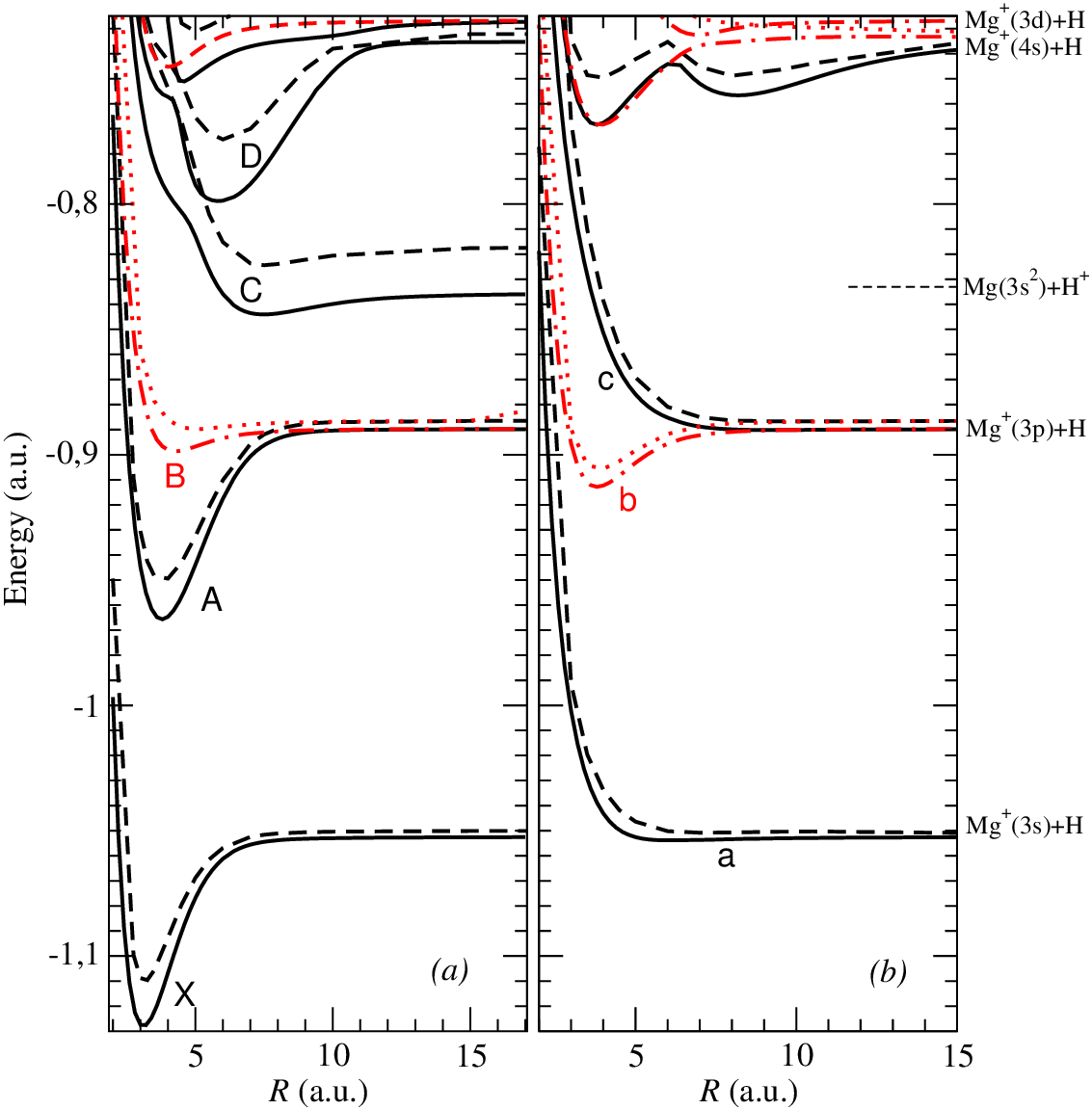}
\end{center}
\caption{(a) Singlet and (b) triplet potential curves MgH$^+$ computed in the present work (full lines: $\Sigma^+$ symmetry; dotted-dashed lines: $\Pi$ symmetry) compared to those of ref.\cite{garcia-madronal1992} (dashed lines: $\Sigma^+$ symmetry; dotted lines: $\Pi$ symmetry). The lowest states are labeled by their standard spectroscopic notation: $X, A, B, C, D$ for $(1)\,^1\Sigma^+ (3s)$, $(2)\,^1\Sigma^+ (3p)$, $(1)\,^1\Pi (3p)$, $(3)\,^1\Sigma^+ (3s^2)$, $(1)\,^1\Sigma^+ (4s)$, respectively, and $a, b, c$ for $(1)\,^3\Sigma^+ (3s)$, $(1)\,^3\Pi (3p)$, $(2)\,^3\Sigma^+ (3p)$, respectively.}
\label{fig:pot_garcia}
\end{figure}

The potential curves for the lowest singlet and triplet states of MgH$^+$, computed with the combination of the \{Mg$^{(C)}$)\} and \{H$^{(A)}$)\} basis sets, are displayed in Figure \ref{fig:pot_garcia}, compared those of ref.\cite{garcia-madronal1992} which are, to our knowledge, the only ones whose numerical data is available in the literature. As discussed in the previous section, this comparison is presented from the point of view of binding energies. The discrepancies reported in Table \ref{tab:atom} on atomic energies are easily identified at large distances. It is not surprising that well depths are smaller in ref.\cite{garcia-madronal1992} than in the present work, due to the limited basis set size. This is particularly true for the $B\,^1\Pi$ and the $a\,^3\Sigma$ states, whose potential wells found in ref.\cite{garcia-madronal1992} are hardly visible at the scale of the figure. The splitting of the large avoided crossing between the $C$ and the $A$ $^1\Sigma^+$ states, which is responsible for charge exchange phenomena at high collision energies \cite{morgan1979,dutta1997}, is reduced according to our calculations, which could then modify the conclusions for the collision experiments.

\begin{figure}[htb]
\begin{center}
\includegraphics[width=0.5\textwidth]{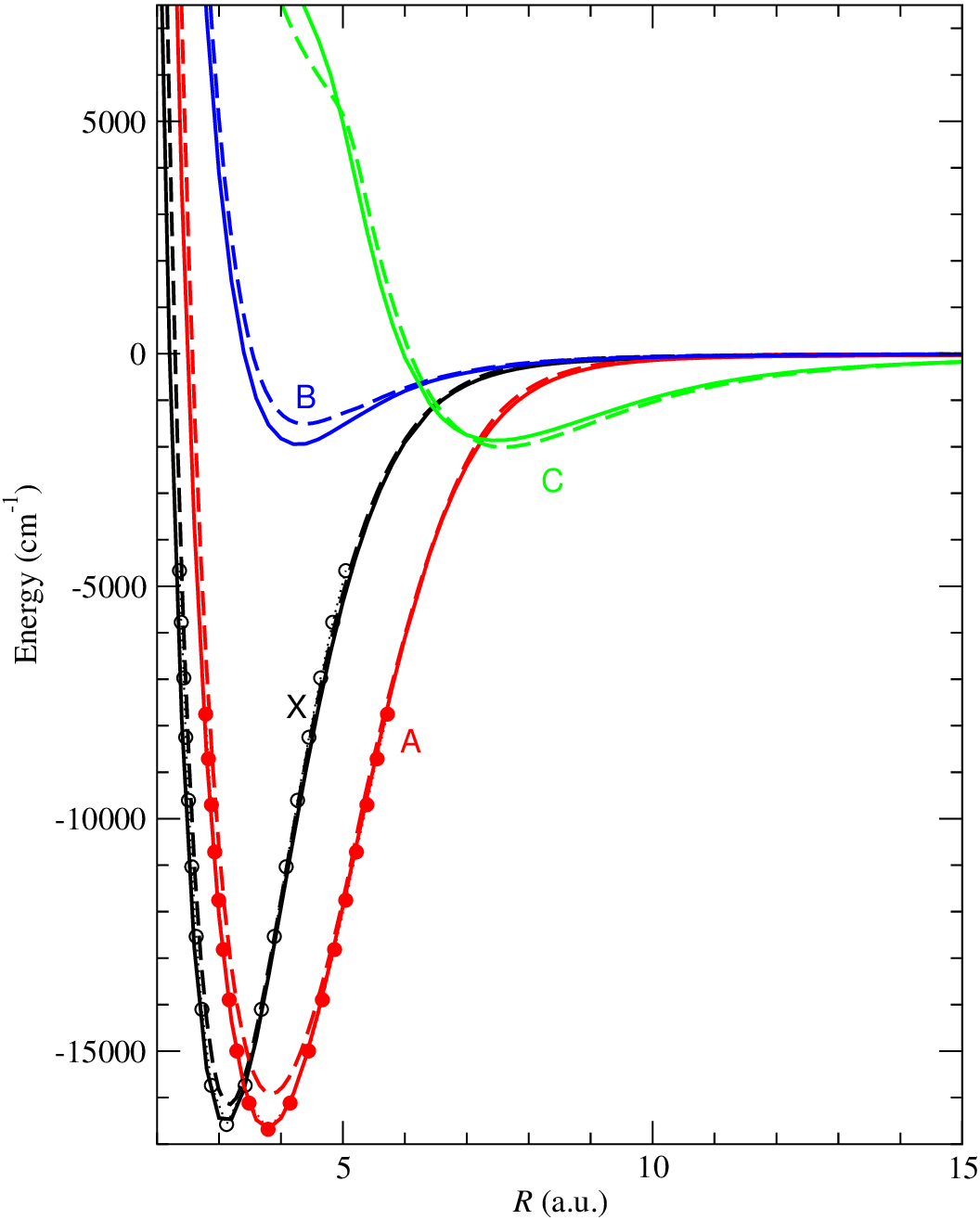}
\end{center}
\caption{Potential curves of the four lowest singlet states of MgH$^+$, using the same spectroscopic notation than in Fig.\ref{fig:pot_garcia}. Full lines: present calculations. Dashed lines: ref.\cite{vogelius2005,jorgensen2005}, at the highest level of theory reported there. All curves are shifted to the same origin, for better visibility. The experimental RKR curves for the $X$ and $A$ states from ref.\cite{balfour1972} are drawn with circles. They are shifted to the bottom of the wells calculated in the present work.}
\label{fig:pot_vogelius}
\end{figure}

In Fig.\ref{fig:pot_vogelius}, we present the comparison of the four lowest singlet potential curves with the recent results of ref.\cite{vogelius2005,jorgensen2005}. In this latter reference, the authors have shifted their potential curves one by one to the experimental value of their corresponding dissociation limit, hoping to minimize the effect of the shift on their computed atomic energies in their modeling of the MgH$^+$ photodissociation. Therefore we have drawn all potential curves relative to the same origin of energy, in order to emphasize on the differences in the shape and depth of the potential curves. The overall agreement between our ECP+CPP+FCI approach and the all-electron MOLPRO calculations is quite satisfactory, especially for the $C$ state, while the well depth of the $X$, $A$, and $B$ states is slightly less pronounced. The experimental potential curve extracted in ref.\cite{balfour1972} from spectroscopic measurements concerns only the low part of the well of the $X$ and $A$ states, and we have drawn them in order to make them match the bottom of the present curves. The agreement in the shape of our potential curves with those of the RKR curves is excellent.

While complete sets of data for potential curves are quite scarce, several authors reported the main spectroscopic constants (equilibrium distance $R_e$, potential well depth $D_e$, and harmonic constant $\omega_e$) for the lowest singlet and triplet molecular states, which are summarized in Tables \ref{tab:constants_1} and \ref{tab:constants_3}, respectively. The main influence of the choice of the basis set combination is visible of the potential well depth, and is only weak on the harmonic constant, while the equilibrium distance is almost stable. Our recommended values would be those yielded by the combination of the  \{Mg$^{(C)}$)\} and \{H$^{(A)}$)\} basis sets, while the \{H$^{(A)}$)\} set still looks reasonable to use. Our results for the ground state are in good agreement with the reported experimental data, except with the one of ref.\cite{dalleska1993}. The situation is less clear for the $A$ state, as we refer to experimental data which are quoted in ref.\cite{numrich1975}, without the original source. Moreover, Balfour \cite{balfour1972} only provided rough estimates of the depth of the potential curves, as his spectroscopic study did not cover the entire potential well. A broad dispersion can be seen in the other theoretical results (apart those of refs.\cite{garcia-madronal1992,jorgensen2005} already discussed above), and our results are in agreement with those of refs.\cite{canuto1993,monteiro1988}. Finally, as it could be expected from the discussion above, the potential well depth of our triplet states is generally larger than the ones available elsewhere \cite{garcia-madronal1992,monteiro1988}. Note that the $c$ state probably has at most a single vibrational level.

\begin{table}[htb]
\begin{tabular}{|c|ccc|ccc|ccc|ccc|} \hline
ref.&\multicolumn{3}{|c|}{$X\,^1\Sigma^+$}&\multicolumn{3}{|c|}{$A\,^1\Sigma^+$}& \multicolumn{3}{|c|}{$B\,^1\Pi$}&\multicolumn{3}{|c|}{$C\,^1\Sigma^+$} \\
    &$R_e$& $D_e$& $\omega_e$&$R_e$& $D_e$& $\omega_e$&$R_e$& $D_e$& $\omega_e$&$R_e$& $D_e$& $\omega_e$  \\ \hline
(a)                           &3.135&15801&1599&3.79 &16500&1143&4.3 &1936&577& &&\\
(b)                           &3.135&15828&1588&3.79 &16913&1130&4.28&1932&588&7.48&1850&325\\
(c)                           &3.095&16553&1599&3.79 &16656&1125&4.28&1943&588&7.48&1860&319\\
(d)                           &3.11 &16501&    &3.79 &16602&    &4.28&1901&580& &&\\
(e)                           &3.11 &16376&    &3.79 &16395&    &4.26&1837&696& &&\\ \hline
\cite{numrich1975}            &3.439&7980 &&4.175&17000&&         & & &&&\\
\cite{rosmus1977}             &3.12 &16130 &1703&&&&&&&&&\\
\cite{olson1979}              &3.163&15600&&3.844&15600&&4.408&1450& & &&\\
\cite{fuentealba1987}         &3.08 &17420&1863&&&&&&&&&\\
\cite{monteiro1988}           &3.15 &16400&&3.86 &15700&&4.29 &1630& & &&\\
\cite{garcia-madronal1992}    &3.17 &13100&&3.82 &13900&&5.1  &720 & & &&\\
\cite{canuto1993}             &3.13 &14945&1682&&&&&&&&&\\
\cite{dutta1997}              &3.118&14300&&3.718&14110&&4.339&1360& & &&\\
\cite{jorgensen2005}          &3.137&16140&&3.84 &15920&&4.38 &1510& & &&\\ \hline
\cite{huber1979}                  &3.12 &16780&1699&&&&&&&&&\\
\cite{numrich1974}            &3.116&16900&&3.79 &16200&&         & & &&&\\
\cite{dalleska1993}           &     &15650 $\pm 500$&&&&&&&&&&\\
\cite{balfour1972}            &3.122&$\approx 17000$&1699&3.780&$\approx 18000$&1135&&&&&& \\ \hline
\end{tabular}
\caption {Equilibrium distance $R_e$ (in a.u.), potential well depth $D_e$ (in cm$^{-1}$), and harmonic constants $\omega_e$ for the $X\,^1\Sigma^+$, $A\,^1\Sigma^+$, $C\,^1\Sigma^+$, and $B\,^1\Pi$  states of MgH$^+$, computed in the present work and compared to values available in the literature. Calculations with various combinations of basis sets are presented in the uppermost rows: (a) \{Mg$^{(A)}$)\}+\{H$^{(A)}$)\}, (b) \{Mg$^{(B)}$)\}+\{H$^{(A)}$)\}, (c) \{Mg$^{(C)}$)\}+\{H$^{(A)}$)\}, (d) \{Mg$^{(C)}$)\}+\{H$^{(B)}$)\}, (e) \{Mg$^{(C)}$)\}+\{H$^{(C)}$)\}. Values displayed in the middle rows are from theory, and from experiment in the lowest rows.}
\label{tab:constants_1}
\end{table}

\begin{table}[htb]
\begin{tabular}{|c|ccc|ccc|ccc|} \hline
ref.&\multicolumn{3}{|c|}{$a\,^3\Sigma^+$}& \multicolumn{3}{|c|}{$c\,^3\Sigma^+$}& \multicolumn{3}{|c|}{$b\,^3\Pi$} \\
    &$R_e$& $D_e$& $\omega_e$&$R_e$& $D_e$& $\omega_e$&$R_e$& $D_e$& $\omega_e$  \\ \hline
(a)                       &6.97&150&125&3.81&4976&904&9.16&151&61\\
(b)                       &6.94&197&122&3.81&4980&896&9.22&78 &66\\
(c)                       &6.06&281&129&3.79&5057&915&9.27&79 &70\\
(d)                       &6.9 &206&&3.82&5008&&9.42&57 &\\
(e)                       &6.82&131&&3.82&4840&&    &&\\ \hline
\cite{monteiro1988}       &7.25&130&&3.83&4660&&9.9 &45 &\\
\cite{garcia-madronal1992}&7.4 &180&&3.9 &4250&&9.5 &60 &\\ \hline
\end{tabular}
\caption {Equilibrium distance $R_e$ (in a.u.), potential well depth $D_e$ (in cm$^{-1}$), and harmonic constants $\omega_e$ for the $a\,^3\Sigma^+$, $c\,^3\Sigma^+$, and $b\,^3\Pi$  states of MgH$^+$, computed in the present work, and compared to values available in the literature. The various displayed cases are labeled in the same way than in Fig.\ref{tab:constants_1}. }
\label{tab:constants_3}
\end{table}

\section{Other properties: Transition and permanent dipole moments, polarizability}
\label{sec:other}

We have computed the electronic transition and permanent dipole moments, and static dipole polarizability, as functions of the internuclear distance, using the various basis set combinations defined in the previous section. As in our previous works, we checked that the difference among the calculations do not exceed about 1\% among each other, in the vicinity of their maximal value. In the figures, we only reported values coming from the combination \{Mg$^{(C)}$)\}+\{H$^{(A)}$)\}.

\begin{figure}[htb]
\begin{center}
\includegraphics[width=0.6\textwidth]{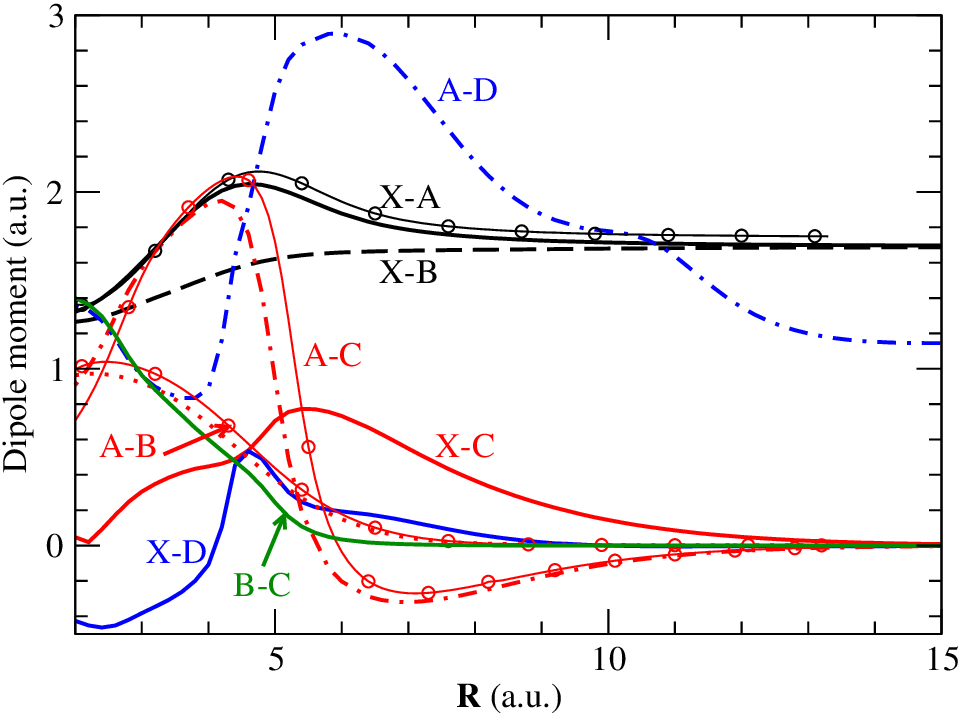}
\end{center}
\caption{Transition dipole moments for selected states of MgH$^+$. Transitions are labeled using the same spectroscopic notation than in Fig.\ref{fig:pot_garcia}. Lines without symbol correspond to the present calculations, while those marked with circles are from ref.\cite{vogelius2005,jorgensen2005}, computed at their highest level of theory .}
\label{fig:trdip}
\end{figure}

Figure \ref{fig:trdip} show the dipole moment functions for all the possible transitions between the five lowest singlet molecular states. Our results are found in good agreement the data from refs.\cite{vogelius2005,jorgensen2005} for the $X-A$, $A-B$, and $A-C$ transitions, which is a strong argument in favor of the quality of both calculations, which proceed along different lines \footnote{In ref.\cite{jorgensen2005}, we changed the sign of the $A-C$ function beyond 6~a.u., as the oscillation shown there actually corresponds to such a sign change.}. Apart from these transitions, the other functions are given for the first time, to our knowledge. The $A-C$ transition plays the main role in the photodissociation model of ref.\cite{jorgensen2005}. The change of sign in the $A-C$ function around 6~a.u. is due to the change of character of the $C$ state due to the avoided crossing with the $A$ state. This function rapidly drops down to zero at large distances, as expected for a transition between states of the Mg$^+$+H pair and of the Mg+H$^+$ pair. The maximum of this function around 5~a.u. coincides with the avoided crossing between the $C$ and $D$ excited states, just like in the $X-D$ function. We also note the local change of the $A-D$ function around 12~a.u. due to the avoided crossing of the $D$ state with an upper excited state. It is likely to expect that the efficient excitation of the $A$ state into the $D$ state, even for high-lying vibrational levels (i.e. with a wave function extending towards large distances) would lead to another photodissociation channel  with charge exchange into the $C$ state, through predissociation across the avoided crossing between the $D$ and $C$ states.

The permanent dipole moment of heteronuclear molecular ions is most often tedious to derive from standard electronic structure calculations, as its calculated value depends on the choice of the origin of coordinates. A straightforward way to overcome this difficulty is to compute the potential energy of the molecular ion placed in a perturbative static electric field with a constant amplitude and direction. This corresponds to the well-known finite-field method suggested by Cohen many years ago \cite{cohen1965}. By varying the strength of the electric field at each distance, the potential energy evolves as a parabola, yielding the permanent dipole moment and the static dipole polarizability as the linear and quadratic coefficients \cite{deiglmayr2008}. If the $z$ axis is chosen along the internuclear axis in the molecule-fixed reference frame ($x$,$y$,$z$), they are two independent components of the molecular polarizability tensor, i.e., the parallel component $\alpha_{\parallel}\equiv \alpha _{zz}$ and the perpendicular one $\alpha_{\perp}\equiv \alpha_{xx} = \alpha_{yy}$. Two related quantities are usually defined: the average polarizability  $\alpha=(\alpha _{zz} +2\alpha_{xx})/3$ and the polarizability anisotropy $\gamma = \alpha _{zz}- \alpha _{xx}$.

\begin{figure}[h]
\begin{center}
\includegraphics[width=0.8\textwidth]{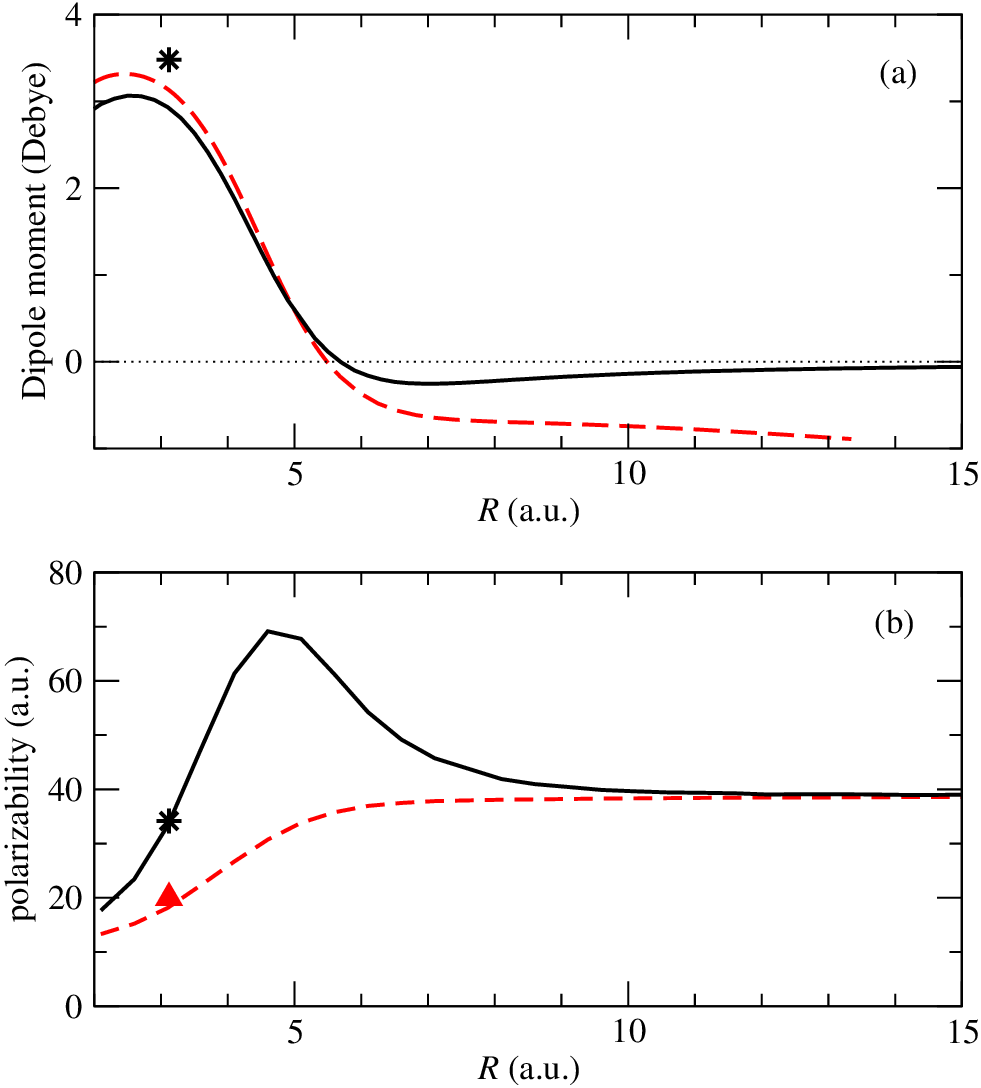}
\end{center}
\caption{(a) Permanent dipole moment of the MgH$^+$ ground state computed in the present work (full line) compared to the one of ref.ref.\cite{vogelius2005,jorgensen2005} (dashed line)computed at their highest level of theory. (b) Parallel (full line) and perpendicular (dashed line) dipole polarizability of the MgH$^+$ ground state computed in the present work. Theoretical values reported in ref.\cite{sadlej1991} at the equilibrium distance (3.12$a_0$) are indicated by stars and triangle.}
\label{fig:polpermdip}
\end{figure}

Our results for the ground state are shown in Figure \ref{fig:polpermdip}, compared to the very few other available theoretical data. The permanent dipole moment is in satisfactory agreement with the one of ref.\cite{jorgensen2005} computed at their highest level of theory, both for the position and amplitude of its maximum and for the position of the change of sign. However we note that the asymptotic value obtained in ref.\cite{jorgensen2005} seems to diverge from a vanishing value. Both components of the static dipole polarizabilities exhibit an $R$-variation similar to the one obtained for alkali dimers \cite{deiglmayr2008}, alkali hydrides \cite{aymar2009}, or for the hydrogen molecule \cite{kolos1967}. In all these systems, the parallel polarizabilities has a maximum at a distance around 1.3 to 1.5 times the equilibrium distance $R_e$ of the system, while the perpendicular components always have a smaller magnitude than $\alpha_{\parallel}$, and monotonically increase towards the asymptotic limit. For all the quantities of Figure \ref{fig:polpermdip}, our results are also in agreement with the data reported at the equilibrium distance by Sadlej and Urban  \cite{sadlej1991} using polarized basis sets and many-body perturbation and coupled-cluster theories combined with finite field method.
At large distances, the polarizability components converge toward  the sum of the atomic values $\alpha_{at}$ of the two constituents.  We found an asymptotic value (39.9~a.u.) which is slightly smaller than the sum of the atomic polarizabilities for H (4.5~a.u. \cite{landau1967}) and for Mg$^+$ (38.84~a.u.) \cite{sadlej1991}.


\section{Conclusion}
In the present paper, we performed an accurate investigation of the electronic structure of the MgH$^+$ ion, including potential curves for singlet and triplet states, transition and permanent dipole moments, and static polarizabilities, as functions of the internuclear distances. Our work extends the recent similar study of ref.\cite{jorgensen2005} using an all-electron approach based on the MOLPRO package, restricted to the lowest singlet states. While minor differences occurred, both approaches are in good agreement, assessing their validity and their good accuracy. By computing higher electronic states, our study suggests that another MgH$^+$ photodissociation path could be worthwhile to try in the framework of the trapped Mg$+$ experiments involving collisions with H$_2$, resulting into the formation of trapped MgH$^+$ molecular ions. Our study represents the necessary initial step towards the description of the quantum dynamics of the Mg$^+ (3p)$+H$_2$ reactivity, which has been recently probed to exhibit a strong isotopic effect \cite{staanum2008}.

\section*{Acknowledgments}
R. G. acknowledges generous support from ANR ({\it Agence Nationale de la Recherche}) under the project CORYMOL, and of IFRAF ({\it Institut Francilien de Recherches sur les Atomes Froids}). M. S. acknowledges support from Universit\'e Paris-Sud 11 for his 3-month stay at LAC. We are indebted to M. Drewsen, S. J\o rgensen, R. Kosloff, P. Staanum, for providing us with the {\it ab initio} data used in I. Vogelius' PhD thesis, as well as D. G. Truhlar for providing us with the data of R. W. Numrich's PhD thesis.

\section*{References}


\providecommand{\newblock}{}

\newpage

\newpage

\newpage

\newpage

\end{document}